\documentclass[12pt]{article} % use larger type; default would be 10pt
\usepackage[toc,page]{appendix}
\usepackage[utf8]{inputenc} % set input encoding (not needed with XeLaTeX)
\usepackage{cite}
\usepackage{color}
\usepackage{epsfig}
\usepackage{graphicx}
\usepackage{subfigure}
\usepackage{ulem}
\usepackage{amsfonts}
\usepackage{epsfig,bm}
\usepackage{graphicx}
\usepackage{geometry} 
\usepackage{graphicx} 
\usepackage{booktabs} 
\usepackage{array} 
\usepackage{paralist} 
\usepackage{verbatim} 
\usepackage{subfig} 
\usepackage{fancyhdr} 
\usepackage{sectsty}
\allsectionsfont{\sffamily\mdseries\upshape} 
\usepackage[nottoc,notlof,notlot]{tocbibind} 
\usepackage[titles,subfigure]{tocloft}

\usepackage{amsfonts}

\title{Thermodynamic Geometry of Yang-Mills Vacua}
\vspace{-0.9cm}
\date{Stefano Bellucci$^{a}$\thanks{\noindent bellucci@lnf.infn.it} 
and Bhupendra Nath Tiwari$^{a, b, c}$\thanks{\noindent bntiwari.iitk@gmail.com}\\
\vspace{0.3cm}
$^{a}$INFN-Laboratori Nazionali di Frascati,\\ Via E. Fermi 40, 00044 Frascati, Italy,\\ 
\vspace{0.2cm}
$^{b}$Amity School of Applied Sciences,\\  
Amity University Haryana, Gurgaon 122413, India\\ 
and\\
$^{c}$University for Information Science and Technology,\\ St. Paul The Apotle, Ohrid 6000, Republic of Macedonia}
\begin{document}
\maketitle
We study vacuum fluctuation properties of an ensemble of $SU(N)$ gauge theory configurations, in the limit of large number of colors, \textit{viz.} $N_c \rightarrow \infty$, and explore statistical nature of the topological susceptibility by analyzing its critical behavior at a nonzero vacuum parameter $\theta$ and temperature $T$. We find that the system undergoes a vacuum phase transition at the chiral symmetry restoration temperature as well as at an absolute value of $\theta$. On the other hand, the long range correlation length solely depends on $\theta$ for the theories having critical exponent $e=2$ or $T=T_d+1$, where $T_d$ is the decoherence temperature. Further, it is worth noticing that the unit critical exponent vacuum configuration corresponds to a noninteracting statistical basis pertaining to a constant mass of $\eta^{\prime}$. 

\vspace{0.50 cm}

\textbf{Key Words:} Thermodynamic Geometry, Fluctuation Theory, Phase Transitions, Topological Susceptibility, QCD, Large $N$ Gauge Theory.

\newpage
\section{Introduction}
QCD as a Yang-Mills gauge theory has opened interesting avenues in understanding the strong nuclear processes and decay reactions \cite{Sakurai, QCD, Witten79}. Wherefore, the stability of nuclear particles, viz. the determination of the lifetime of a particle or decay rate of a given nuclear reaction, is one of the most interesting research issue since the proposition of modern gauge field theory \cite{gauge}. As particles are considered as certain excited states, thus we are going to examine the stability of vacuum configurations under variations of the parameters. Such an intrinsic principle has been highly successful when applied in studying both the particle properties and decay rate based applications \cite{QCD}. This analysis follows further because of the fact that it is easy in its implementation and it requires only finitely many parameter tuning. Indeed, the  fluctuation geometric characterizations under question are the best alternative to offer stability properties of QCD vacuum, as a statistical system. This follows from the fact that the intrinsic geometric consideration takes an account of configuration uncertainties and generally allows for a clear-cut determination of regular and singular domains of the vacuum parameters. 

Motivated from Witten's work on CP breaking and duality principle between large $N$ gauge theories and string theory on certain spacetimes in the realm of wrapped six-branes describing a set of adjacent vacua which are separated by domain walls \cite{Witten98}, the above observation is further supported from the fact that a class of stringy vacua remains insensitive to temperature variations as well as certain parametric deviations, see for instance extremal black hole configurations in string theory \cite{SV}. Fluctuation theory \cite{rupp1, rupp2,Ruppeiner3,Ruppeiner4,McMillan,Simeoni,widom} based parametric modeling appears one of the best methods to explicate the stability of a given statistical system. This is because of the fact that the intrinsic geometric approach enables us in determining both the local and global stability characteristics of a chosen configuration. 

The parameters concerning the parity odd bubbles \cite{kharjev, Witten79} are the vacuum angle $\theta$and temperature $T$ of the undermining configuration. Therefore, the phase determination of an ensemble of QCD vacua, viz. the local and global characterizations depend on the fluctuation range of $\{\theta, T \}$. In this line of research, Ref.\cite{kharjev} leads us to provide an appropriate mathematical expression and an intrinsic parameterization to the free energy of parity odd bubble configuration. From this outset, the present paper focuses on the determination of stable embedding functions under intrinsic geometric fluctuations concerning the minimization of vacuum uncertainties.
Indeed, for an arbitrary non-equilibrium gauge or field theory configuration, the above parametric geometric hypothesis can be carried forward in terms of the associated effective potential parameterized via moduli fields. Such an intrinsic stability analysis of gauge theory vacuum configurations we leave open for a future research. 
 
In the light of our geometric model, important issues pertaining to the statistical stability of parity odd bubbles are determined as an undermining embedding \cite{bntenfun} via the fluctuation theory analysis \cite{rupp1, rupp2} of free energy and effective potential of a given large $N$ gauge theory configuration \cite{Witten79}. The parametric determination of the free energy  of such a configuration is based on an extension of the scattering theory and anomaly cancellation \cite{Witten79,veneziano}. A novel formulation is consequently made realizable in stabilizing parity odd bubble ensembles via the present consideration. Namely, our intrinsic geometric method as outlined in the sequel is very generic in its own right and it opens various possible directions for further investigations. One of the fundamental issues on which we focus in this paper is an appropriate stability determination of an ensemble of QCD vacua. To do so, one traditionally assumes that one of the parameter of the vacuum lies in subset of parametric family of the physical system, \textit{i.e.}, the free energy \cite{Witten79} depends on the vacuum parameter $\theta$. For instance, see \cite{Witten98} concerning spontaneous breaking of CP symmetry and large $N$ gauge theories in four-dimensional spacetimes.

Ref \cite{kharjev} offered an extension of fundamental principles such as Witten's scattering theory analysis \cite{Witten79} and Veneziano's anomaly consideration \cite{veneziano} for the associated mesons concerning parity odd bubbles. From a closer perspective, such an analysis of the large $N$ gauge theory takes an account of the free energy determination. Specifically, it allows a straightforward relation of vacuum settings in terms of the associated effective field theory model parameters. Notice further that the consideration of vacuum fluctuations is insensitive to supersymmetry and gauge coupling deviations, and the output is approximately equivalent to a large $N$ pure $SU(N)$ gauge theory ensemble. In this paper, in order to explore stability of such ensembles, we focus on the designing of a stable parity odd bubble vacuum region from the perspective of the intrinsic Riemannian geometry.

Based on the covariant geometric consideration, we offer stability analysis of parity odd bubble ensembles in general. The fluctuation theory approach - - based on the large $N$ free energy evaluation \cite{kharjev} with an incorporation of nonzero temperature and anomalous couplings of mesons - - enabling one in providing improved stability structures of bounded particles and associated decays, is presented in this paper. Graphical representations of local and global parametric fluctuations are shown for the limiting large $N$ configuration. We anticipate that the  geometric framework is particularly well suited for a variety of practical applications. It turns out that the mathematical designing procedure thus illustrated can further be extended for arbitrary vacuum ensemble. Indeed, there exists a broad class of perspective investigations as far as this representation is concerned. This is one of the real bestows for stabilization of parity odd bubble vacuum ensembles. The present analysis of parity odd bubbles can in turn be pushed further in explicating intrinsic nature of LHC data \cite{LHC} concerning the standard model of particle physics and associated background disturbances arising from vacuum fluctuations.

\`{A}priori, we have considered the above intrinsic geometric analysis for a class of equilibrium statistical configurations involving extremal and non-extremal black holes in string theory \cite{9601029v2,9411187v3,9504147v2,0409148v2,9707203v1,0507014v1, 0502157v4,0505122v2} and $M$-theory \cite{0209114,0401129,0408106,0408122}, and shown that such ensembles possess rich intrinsic geometric structures, see for instance \cite{0606084v1,SST,0801.4087v1,BNTBull, BNTBull08,BNTBullcorr, RotBH,BSBR}. Concerning the physics of black holes, black strings, black rings and black branes in general, there has been much well focused attention on equilibrium statistical fluctuation perspective of string theory vacua \cite{0606084v1,SST,0801.4087v1,BNTBull,BNTBull08,BNTBullcorr,RotBH,BSBR}, in order to explicate the nature of undermining parametric pair correlation functions and stability of the associated string theory solution containing a set of branes and antibranes. Indeed, there have earlier been several general notions analyzed in the condensed matter physics \cite{RuppeinerRMP,RuppeinerA20, RuppeinerPRL,RuppeinerA27,RuppeinerA41,RuppeinerPRD78}, as well. Based on the above motivations, we focus our attention on large $N$ gauge theory configurations as mentioned above with given equilibrium parameters $\{\theta, T\}$. Thereby, we examine possible parametric local pair correlations and the associated global correlation length undermining the parity odd hot QCD bubbles. In nutshell, the present research when analyzed in terms of system parameters offers an intriguing dimension to fluctuating vacuum ensembles.

The questions that we attempt to answer in this paper are: (i) under what constraints a considered vacuum ensemble is (un)stable?, and (ii) how its parametric vacuum correlation functions scale in terms of the chosen fluctuating parity odd bubble parameters?  With such a definite covariant geometric description of a consistent equilibrium statistical system, we can indeed determine the complete set of non-trivial local correlation relations of gauge theory vacuum configurations. It is worth mentioning further that similar considerations hold for the four dimensional black hole solutions in general relativity \cite{gr-qc/0601119v1,gr-qc/0512035v1, gr-qc/0304015v1,0510139v3}, and equilibrium (un)attractor black holes \cite{9508072v3,9602111v3,new1,new2,0702019v1, 0805.1310}, as well. Moreover, based on the Legendre transformed finite temperature chemical configurations \cite{Weinhold1, Weinhold2}, we have explored the above thermodynamic type stability analysis under parametric fluctuations of the 2- and 3- flavor hot QCD systems \cite{BNTSBVC, bntsbvc1, bntsbvc2, bntsbvc3}.  As per the above outset, we herewith pronounce that the consideration of differential geometry plays a vital role in studying stability properties of both the gauge theory and field theory configurations.

In this paper, we offer the local and global stability criteria of an ensemble of large $N$ parity odd bubbles under fluctuations of the vacuum angle $\theta$ and QCD temperature $T$. As the function of $\{\theta, T\}$, the free energy $F(\theta, T)$ under consideration \cite{kharjev} is depicted in the Fig.(\ref{freeE}). The stability characteristics is demonstrated for a suitable set of parameters $\{\theta, T\}$ rendering a specific parameterization as an undermining embedding \cite{bntenfun}. Namely, as for as the parity odd bubbles are concerned, the free energy offers such an intrinsic characterization of vacuum fluctuations via the following real embedding: $(\theta, T) \hookrightarrow  F(\theta, T)$. From the general $SU(N)$ gauge theory parametrization equation in the large $N$ limit \cite{kharjev}, the stability of component diagram is brought out into an attention. It is worth mentioning that this formulation is made possible for the first time towards the intrinsic geometric stability analysis of an ensemble of parity odd bubbles. The proposed methodology of intrinsic geometry is rather very generic and different Legendre transformed versions of the  present consideration are possible. We leave an intrinsic classification of such transformations open for a future research. In this paper, we focus on the metric geometries originating from the energy representations.

By employing standard notions of the intrinsic Riemannian geometry, the rest of  sections are devoted in determining local and global stability properties of an ensemble of fluctuating parity odd bubble configurations. In section 2, we give review of Yang-Mills theory, whereby the framework of the embedding geometry via the formation of an admissible energy surface. The section 3 is devoted to examine fundamental intrinsic geometric setup in the light of the fluctuation theory. Hereby, for an ensemble of fluctuating YM vacua, we highlight the statistical correlation outcomes as emerging via our free energy based embedding geometry. In section 4 we examine stability properties arising from fluctuations of the free energy $F(\theta, T)$ as a function of $\{\theta, T\}$ concerning an undermining ensemble of parity odd bubble configurations. Finally, in section 5, we offer our fluctuation geometric conclusions and perspective directions for future research.
\section{Review of the Model}
Let us begin by considering an ensemble of $SU(N)$ gauge theory in the large color limit $N \rightarrow \infty$. In the following, we exploit the large $N$ expansion \cite{Hooft} to investigate the stability behavior of statistical ensembles at a nonzero temperature, especially while examining local and global statistical correlation functions concerning the corresponding topological susceptibility. Namely, by invoking joint fluctuations of the vacuum parameter $\theta$ and temperature $T$, we wish examining both the nature and order of deconfining phase transitions. Before doing so, we offer a brief recap of the Yang-Mills gauge theory \cite{Sakurai, QCD, Witten79}, fluctuation theory \cite{rupp1, rupp2}, and an outline of fluctuating gauge theory vacuum ensembles in the realm of embedding theory.
\subsection{Recaps on Yang-Mills Theory}
As mentioned in the introduction, recall that the deconfining temperature is defined as the temperature $T_{\chi}$ corresponding to the restoration of the chiral symmetry. Secondly, for a given YM theory \cite{Sakurai, QCD, Witten79, kharjev}, it is well known that the topological charge density can be expressed as the following total derivative
\begin{equation}
Q(x) = \frac{g^2}{32 \pi^2} tr( G_{\alpha \beta} \tilde{G}^{\alpha \beta}) = \partial_{\alpha} K^{\alpha},
\end{equation}
where $ K^{\alpha}$ is a gauge dependent current.
 
In terms of the above charge $Q(x)$, the topological susceptibility is defined as the two point function of $Q$, viz. we have  
\begin{equation}
\lambda_{YM}= \frac{\partial^2 F( \theta, T)}{\partial \theta \partial T} = \int d^4 x <Q(x)\ Q(0)>,
\end{equation}
where $\theta$ is conjugate to the topological charge density $Q$ and the zero temperature free energy $F(\theta, 0) = E(\theta)$ is termed as the $\theta$-dependent vacuum energy.
Since $Q(x)$ is a total space-time derivative of the gauge dependent current $ K^{\alpha}$, the topological susceptibility $\lambda_{YM}$ receives only instantonic contributions order by order, in the $\frac{1}{N}$-expansion \cite{Hooft, Witten1byN}. In this case, it follows that the topological susceptibility satisfies the following leading order behavior 
\begin{equation}
\lambda_{YM}(T)= \lambda_{YM}(0) e^{-a N}, 
\end{equation}
where $a= \frac{8 \pi^2}{g^2 N}$ for a single instanton contribution. Also see \cite{ref18} for $CP^N$ modes in $(1+1)$ dimensions. At $T=0$, Witten \cite{Witten1byN} suggested that quantum fluctuations generate a nonzero value for the $\lambda_{YM}(0)$, over essentially vanishing semi classical fluctuations to the deconfining phase of the topological susceptibility, viz. one has  $\lambda_{YM}(0) \sim N^0$. Combining the above two phases with the setup of Refs. \cite{Witten1byN, veneziano}, one has the following leading order topological susceptibility 
\begin{eqnarray}
\lambda_{YM}(T) \sim \bigg\lbrace \begin{array}{cc} e^{-a N}; & T\ne 0, \\ N^0; & T = 0 \end{array} 
\end{eqnarray}
as a function of the temperature $T$. In the line of Witten's  scattering theory analysis \cite{Witten1byN} and Veneziano's anomaly cancellation \cite{veneziano}, generalizations to a nonzero temperature configuration \cite{kharjev} have further been considered with anomalous couplings of mesons in literature, for instance see \cite{ref22} for observations concerning radiative decays beyond chiral limit, topological susceptibility and associated Witten-Veneziano mass formula in the measurements of $\eta^{\prime} (\eta) \rightarrow \gamma \gamma$. The above setup results into the following $\theta$ dependent free energy
\begin{equation}
F(\theta, T)= A \ (1+ c \theta^2) (T_d - T)^{2-\alpha}, 
\end{equation}
where $A$ is a known constant and $\alpha$ is critical exponent of the system. In fact, for a given deconfining phase transition temperature $T_d$, Ref. \cite{kharjev} suggests a small negative value of $\alpha$, namely, $\alpha= -0.013$. In the above setup of $SU(N)$ YM- gauge theory, we consider an ensemble of YM vacua, more precisely, we focus on the metastable states parametrized by finitely many (in)equilibrium statistical parameters of the chosen Yang Mills gauge theory configuration.

\subsection{Admissible Energy Surfaces}
  
In the sequel, motivated from fluctuations away from the vacuum fixed point systems in the light of scattering theory, namely, the Witten's $\eta^{\prime}$ mass formula \cite{Witten79,Witten98} and associated complete algebraic saturation of Ward identities in the double limit of Veneziano \cite{veneziano}, we wish studying statistical stability characterizations of the undermining vacuum fluctuations as a function of the vacuum parameters $\{\theta, T\}$. In particular, we focus our attention on  fluctuations of the topological susceptibility vanishing at the critical temperature line $T= T_d$, where $T_d$ is the decoherence temperature of the chosen QCD vacuum system under consideration. 

To discuss the stability properties of such a statistical system, let us consider an ensemble of  meta-stable states acting as regions of nonzero vacuum angle $\theta$ at a given nonzero temperature $T$. Thence, the fluctuation theory setup \cite{rupp1} enables us in examining the statistical stability properties of an ensemble of fluctuating colored $SU(N)$ gauge theory vacuum configurations in general by introducing an admissible intrinsic surface
\begin{equation} \label{admissibleis}
\mathcal{M}_2:= \{ (T, \theta)\ | \ T < T_d,\ \theta \equiv \theta + 2 \pi \}
\end{equation}
Hereby, we focus on the statistical ensembles with sufficiently light quark masses in order to statistically address the signature and production of mesons including the $\eta$ and $\eta^{\prime}$ mesons, as well. This allows us concentrating on the probabilistic admissibility of normally forbidden processes, e.g., $\eta \rightarrow \pi^{0}\pi^{0}$, concerning  the $SU(N)$ gauge theory in the large $N$ limit. Our conclusions are based on Witten's formula pertaining to $U(N)$ problem in the light of $\frac{1}{N}$ expansion, that's, the hadronic phase is cold, viz. the $\eta^{\prime}$ mass vanishes at the critical temperature line $T= T_d$.
\section{Fluctuation Theory Perspective}
In this subsection, we provide fundamental intrinsic geometric setup of the fluctuation theory along with statistical outcomes concerning an ensemble of YM vacua. 
Our consideration is physically based on metastable states, which act like geometric regions of nonzero $\theta$ and $T$, forming a two dimensional intrinsic surface $\mathcal{M}_2$, as above in Eqn.(\ref{admissibleis}). In the sequel, we may consider $\mathcal{M}_2$ as an inner product space of the admissible system of parameters $\{ \theta, T \}$. Namely, it follows that the surface $\mathcal{M}_2$ defines the following map 
\begin{equation} \label{embeddingf}
\{ \theta, T \} \mapsto^F\  F( \theta, T): \mathcal{M}_2 \hookrightarrow  \mathbb{ R}.
\end{equation}
The above embedding involving Riemannian geometry on two dimensional surface of fluctuating parameters $\{ \theta, T \} $ can be explicated as follows. Given a differentiable manifold $\mathcal{M}_2$ of dimension $2$, a Riemannian metric $g$ on it is defined as a family of positive definite inner products $g_p \colon T_p\mathcal{M}_2\times T_p\mathcal{M}_2\longrightarrow \mathbb R$, where $p\in \mathcal{M}_2$ such that the map $p\mapsto g_p(X(p), Y(p))$ gives a smooth function as the real embedding $ \mathcal{M}_2 \hookrightarrow \mathbb{R}$ for any pair of differentiable vector fields $\{X, Y\}$ on $\mathcal{M}_2$. 

Physically, we can view the Riemannian metric $g$ on $\mathcal{M}_2$ as a symmetric $(0,2)$ tensor possessing the positive definiteness property of the metric $ g(X, X) > 0$ for any nonzero tangent vector $X$. For the above real manifold $\mathcal{M}_2$ of dimension $2$, a local coordinate system $\{ x^i\ \vert \ i= 1,2 \}$ on it can be designed in terms of the fluctuation parameters $\{ \theta, T\}$. In a system of local coordinates on the above parametric surface $\mathcal{M}_2$ given by doubled real valued functions $ \{ \theta, T \}$ with the corresponding vector fields $\left\{\frac{\partial}{\partial \theta}, \frac{\partial}{\partial T}\right\}$ on $\mathcal{M}_2$ forming a tangent basis of the tangent manifold $T_p\mathcal{M}_2$ at each point of the parametric surface $\mathcal{M}_2$. At each point $p\in \mathcal{M}_2$, it follows further that the components of the metric tensor can be expressed as the following inner product
\begin{equation}
g_{ij}(p):=g_{p}\Biggl(\left(\frac{\partial }{\partial x^i}\right)_p,\left(\frac{\partial }{\partial x^j}\right)_p\Biggr), i, j = 1, 2.
\end{equation}
with respect to a chosen local coordinate system $(x^1, x^2)$ on the parametric surface $\mathcal{M}_2$ as the vacuum parameters $\{ \theta, T \}$. Equivalently, in terms of the dual cotangent basis $\{dx^1, dx^2 \}$, it follows that the metric tensor for $\mathcal{M}_2$ can be written as
\begin{equation}
 g=\sum_{i,j=1,2}g_{ij}\mathrm d x^i\otimes \mathrm d x^j
\end{equation}
Following the above inner product structure corresponding to the Riemannian metric tensor $g_{ij}$, the fluctuation manifold of the vacuum parameters $ \{ \theta, T \}$ renders us a two dimensional differentiable Riemannian surface $(\mathcal{M}_2, g)$. For standard differential geometric notions such as flow components, correlation components as the elements of the Hessian matrix of a defining embedding function, Christoffel connection functions and Riemannian scalar curvature, there have numerous applications in all domains of the energy scale from condensed matter systems to high energy configurations. For such an explicit evaluation of differential geometric notions in different domain of research, see \cite{0606084v1,SST,0801.4087v1,BNTBull,BNTBull08,BNTBullcorr,RotBH,BSBR} for string theory vacua of branes and antibranes, \cite{RuppeinerRMP,RuppeinerA20,RuppeinerPRL,RuppeinerA27, RuppeinerA41,RuppeinerPRD78} for configurations in condensed matter physics and \cite{BNTSBVC,bntsbvc1, bntsbvc2, bntsbvc3} for hot QCD systems with two and three flavors.

In this paper, we firstly specialize to examine the vacuum temperature fluctuations whose free energy $F( \theta, T)$ we have given in the foregoing subsection. In this setup, we consider a probabilistic interpretation of the parity odd bubble configurations concerning the parity violation, viz. the question whether the $\eta$ particle decays to two pions or instead to the usual three pions \cite{kharjev}. We also propose a global variable, namely, the intrinsic scalar curvature on the fluctuating free energy surface $\mathcal{M}_2$ to measure the statistical interaction properties and long range global correlation length of the undermining ensemble of parity odd bubbles.

For a fixed number of fermion flavors as $N \rightarrow \infty$, it is worth recalling that the large $N$ limit of $SU(N)$ gauge theory is very much gluonic in nature \cite{kharjev} because $N^2$ gluons totally dominate the $N$-quarks. At a large value of $N$, it is well known that the confinement of quarks occurs \cite{Witten80, Hooft81}, thus at a low temperature, we can focus entirely on vacuum correlations of the mesons and glueballs. Hereby, by the term confinement we mean that all the traces of the color indices disappear, and herewith the bound states thus formed are solely characterizable via symmetries of the considered gauge theory model, e.g., spin, parity, etc.

In the sequel, we consider the setup of an arbitrary large $N$ gauge theory in which massless quarks couples to the gluons. As far as the physics at low temperature is concerned, we further know that the chiral symmetry breaks down to a diagonal subgroup of the flavors, see for example \cite{kharjev, veneziano}. Thus, we may assume in standard fashion that the chiral symmetry is restored at the chiral temperature $T_{\chi}$ with $T_{\chi}= T_d$, where $T_d$ is the decoherence temperature. To simplify the above consideration, we may deal with an intrinsic scale of the deconfining transition as to be the same as that of the gluball masses, that's achieved by taking the decoherence temperature limit $ T_d \rightarrow 1$, as $ N \rightarrow \infty $. To investigate ensemble properties of $SU(N)$ gauge theory, we may thus focus on the large $N$ limit of the YM gauge theory without confinement, e.g., the $N$-component vector model with YM coupling $g^2$ such that $g^2N$ remains fixed in the limit of $N \rightarrow \infty$.

Since the gluons dominate the free energy above the deconfining temperature, viz. $T\ge T_d$, thus it follows that the effects of quarks can be neglected in this range of the temperature $T$. The crucial thing which we obtain hereby is the nature of deconfining phase transitions in concerning an ensemble of vacuum $SU(N)$ gauge theories. Namely, we focus on an ensemble of $SU(N)$ metastable states undermining parity odd bubble vacuum phase transitions. This enables us in determining phase transition curves on an intrinsic geometric surface $\mathcal{M}_2$ of the vacuum fluctuation parameters $\{\theta, T\}$ concerning the ensemble of parity odd bubbles. 

As a result, we obtain statistical correlations as a function of the critical exponent under the fluctuations of $(\theta, T)$ . In fact, we find that the global statistical correlation exists, even if we switch off the vacuum parameter $\theta$ and temperature $T$, viz. take the limits: $\theta \rightarrow 0$ and $ T \rightarrow 0$. In this limit, we have a nonzero value of the correlation area $A(\theta, T) \sim R(\theta, T)$, where $R(\theta, T)$ is the scalar curvature on the parametric surface $\mathcal{M}_2$. Namely, in section $3$, we have shown that our statistical analysis of the YM free energy vacuum fluctuations renders the following limiting correlation area
\begin{equation} \label{limitingcur}
\tilde{A} =  lim_{\theta \rightarrow 0,\ T \rightarrow 0} \ A(\theta, T)\ \ \propto \ - \ \frac{T_d^{e-2}}{e-1},
\end{equation}
Herewith, we observe that the $SU(N)$ gauge theory vacuum configuration, even in the limit of zero temperature and zero vacuum parameter $\theta$, corresponds to an interacting statistical ensemble. This follows in particular from a non vanishing value of the correlation length $l=  \sqrt{\tilde{A}}$.
Hereby, we see that the limiting statistical configuration becomes strongly interacting about the unit critical exponent, as $l \rightarrow \infty$ whenever $e \rightarrow 1$. Moreover, we notice that $l$ never vanishes as long as the decoherence temperature $T_d$, undermining the chiral symmetry restoration temperature of metastable states, takes a nonzero finite value. 
\section{Free Energy Fluctuations}
\begin{figure}
\hspace*{1.5cm}
\includegraphics[width=10.0cm,angle=0]{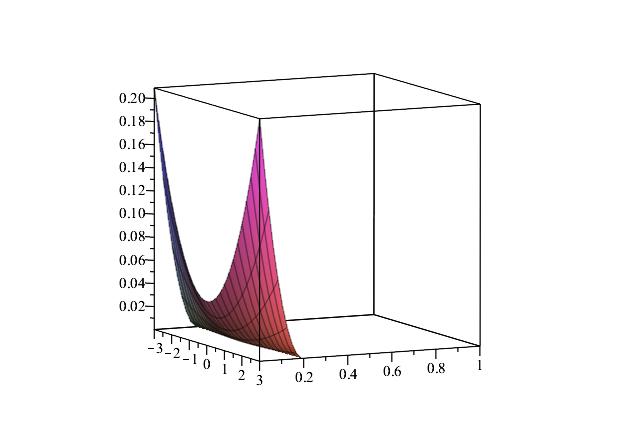}
\caption{The free energy $F$ as the function of vacuum parameter $\theta$
and temperature $T$ describing YM vacuum configuration for a given anomaly factor $c$, decoherence temperature $d$ and critical exponent $e$.}  \label{freeE}
\vspace*{0.6cm}
\end{figure}

The free energy of the present interest is depicted in the Fig.(\ref{freeE}). Although our investigation of the present paper remains valid for any embedding geometric configuration, nevertheless we wish to illustrate it for a class of large N vacua, which are of an immediate interest to us in concerning the fluctuating parity odd bubbles. Here, the parameter $\theta$ is the vacuum angle arising as a translational shift to the total angle of phases. In the subsequent analysis, we invariably denote a locally deconfining QCD phase by the corresponding decoherence temperature as $T_d$or $d$. Moreover, we observe that the stability analysis considered in the framework of intrinsic geometry remains valid for general effective field theory configurations. Therefore, we anticipate further that such an investigation remains consistent for an arbitrary fluctuating gauge theory configuration. 

Given an ensemble of parity odd bubbles, we consider the intrinsic geometric analysis in order to determine the stability properties of an ensemble of large $N$ vacua. The stability criteria under present consideration offer an adept method in ascertaining the required values of system parameters and thereby the undermining local and global stability properties of an ensemble of fluctuating parity odd bubbles. In sequel, we denote the vacuum angle by $x$, temperature as $y$ and the decoherence temperature as $d$. In this setup, the free energy \cite{kharjev} undermining the deconfining phase transition can be represented as per the following expression
\begin{equation} \label{freeenergy}
F(x,y)=\left(1+c x^2\right)(d-y)^{2-e},
\end{equation}
where $c$ is the coefficient of the anomaly and $e$ is the critical exponent. It is worth observing that the free energy $F(x,y)$ as given in Eqn.(\ref{freeenergy}) vanishes identically at the decoherence temperature, viz. $y = d$. Further, $F(x,y)$ becomes independent of the temperature $y$ at the value of critical exponent $e=2$ and it becomes independent of the vacuum angle $x$ at vanishing anomaly contribution $c=0$. At the origin $x = 0$ and $y= 0$ of an undermining real fluctuation surface $\mathcal{M}_2(\mathbb{R})$, we see that the free energy $F(x,y)$ possesses a constant value of $d^{2-e}$.
Herewith, we observe that the flow components of free energy fluctuations are
\begin{eqnarray}
F_x(x,y) &=& 2 c x (d-y)^{2-e}, \newline \nonumber \\ 
F_y(x,y) &=& -(2 - e) (1 + c x^2) (d - y)^{1 - e}
\end{eqnarray}
In this case, we hereby observe that the flow equations defined as $F_x= 0 = F_y$ constitute to the fact that that the vacuum parameter $x$ vanishes identically and temperature $y$ remains fixed at its decoherence value $d$. However, we see that the value of the critical exponent $e$ plays a crucial role in determining the fixed points of the free energy $F(x,y)$. Namely, the choice of $e=2$ makes the first flow equation $F_x=0$ independent of the temperature $y$, while the choice of $e=1$ makes the second flow equation $F_y=0$ independent of $y$. Furthermore, if the vacuum angle is allowed to become an imaginary number, then the value $x= \pm i / \sqrt{c}$ constitutes a complex fixed point of $F(x,y)$. In the sequel of this section, we further explain the possible roles of the critical exponent $e$ while determining the local and global statistical correlations concerning an ensemble of parity odd bubbles.
\begin{figure} 
\hspace*{1.5cm}
\includegraphics[width=8.0cm,angle=0]{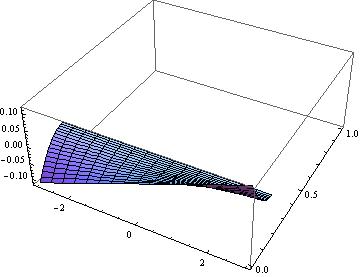}
\caption{The $\theta$ component $F_{\theta}$ of the free energy flow as the function of vacuum parameter $\theta$ and temperature $T$, describing fluctuations of the YM vacuum ensemble for a given anomaly factor $c$, decoherence temperature $d$ and critical exponent $e$.} \label{Ftheta}
\vspace*{0.5cm}
\end{figure}
\begin{figure}
\hspace*{1.5cm}
\includegraphics[width=8.0cm,angle=0]{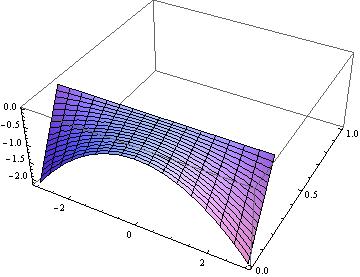}
\caption{The $T$ component  $F_{T}$ of the free energy flow as the function of vacuum parameter $\theta$ and temperature $T$, describing fluctuations of the YM vacuum ensemble for a given anomaly factor $c$, decoherence temperature $d$ and critical exponent $e$.}  \label{Ft}
\vspace*{0.5cm}
\end{figure}
The graphical representations of the flow components $\{ F_{\theta}, F_{T}\}$ are shown in the Figs.(\ref{Ftheta}, \ref{Ft}) with their respective magnitudes of variations in the range $(-0.1, +0.1)$ and $(0.0, -2.0)$. We now wish to describe the intrinsic stability of an ensemble of parity odd bubbles with fluctuating vacuum angle $x$ and temperature $y$. The parametric correlations are described by the Hessian matrix of the underlying free energy $F(x, y)$, defined with a set of desired corrections over a chosen temperature $y$ below the decoherence temperature $d$. Following Eqn.(\ref{freeenergy}), we see that the components of the metric tensor $g$ on the surface $\mathcal{M}_2(\mathbb{R})$ - - defined via the Hessian matrix $Hess(F(x, y))$ of $F(x, y)$ - - reduce as the following expressions:
\begin{eqnarray}
F_{xx} &=& 2 c (d-y)^{2-e}, \newline \nonumber \\ 
F_{xy} &=& -2 c x (2-e) (d-y)^{1-e}, \newline \nonumber \\ 
F_{yy} &=& (1-e) (2-e) \left(1+c x^2\right) (d-y)^{-e}
\end{eqnarray}
In this framework, we notice that the intrinsic geometric nature of parametric vacum pair correlations divulges local notions of fluctuating hot QCD bubbles. Namely, at the decoherance temperature $y=d$, we see that all the local pair correlations $\{ F_{ij}\ \vert \ i, j= x, y \}$ vanish identically for all $x, y \in \mathcal{M}_2(\mathbb{R})$. Whilst, for systems having critical exponent $e=2$, only the local vacuum pair correlation $F_{xx}$ survives with a constant value of $2c$, where $c$ is anomaly factor of the chosen vacuum ensemble. On the other hand, for the systems having unit critical exponents $e=1$, we find that the local temperature pair correlation function $F_{yy}$ further vanishes identically, whilst it is interesting to notice in this case that the local vacuum pair correlation function $F_{xx}$ becomes linear in temperature $y$, and the local cross correlation function $F_{xy}$ modulates linearly in the vacuum angle $x$.

\begin{figure}
\hspace*{1.5cm}
\includegraphics[width=8.0cm,angle=0]{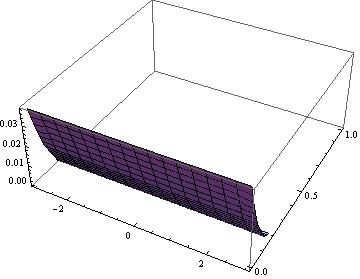}
\caption{The $\theta \theta$ component  $F_{\theta \theta}$ of the free energy fluctuations as the function of vacuum parameter $\theta$ and temperature $T$, describing fluctuations of the YM vacuum ensemble for a given anomaly factor $c$, decoherence temperature $d$ and critical exponent $e$.} \label{Fthetatheta}
\vspace*{0.5cm}
\end{figure}

\begin{figure}
\hspace*{1.5cm}
\includegraphics[width=8.0cm,angle=0]{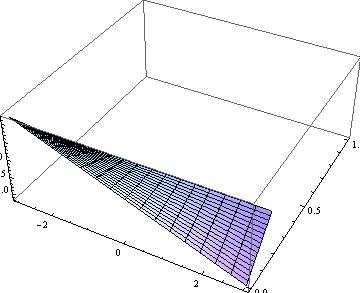}
\caption{The  $\theta T$ component  $F_{\theta T}$ of the free energy fluctuations as the function of vacuum parameter $\theta$
and temperature $T$, describing fluctuations of the YM vacuum ensemble for a given anomaly factor $c$, decoherence temperature $d$ and critical exponent $e$.} \label{Fthetat}
\vspace*{0.5cm}
\end{figure}

\begin{figure}
\hspace*{1.5cm}
\includegraphics[width=8.0cm,angle=0]{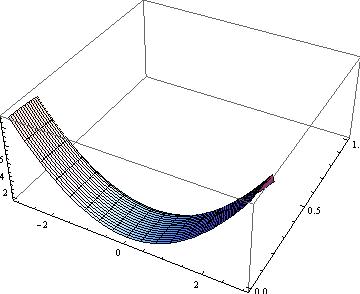}
\caption{The $TT$ component  $F_{TT}$ of the free energy fluctuations as the function of vacuum parameter $\theta$ and temperature $T$, describing fluctuations of the YM vacuum ensemble for a given anomaly factor $c$, decoherence temperature $d$ and critical exponent $e$.} \label{Ftt}
\vspace*{0.5cm}
\end{figure}

\begin{figure}
\hspace*{1.5cm}
\includegraphics[width=8.0cm,angle=0]{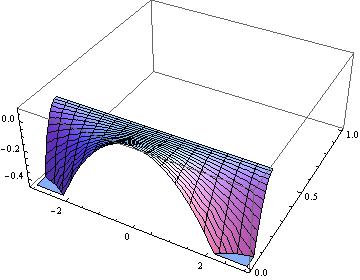}
\caption{The determinant of the metric tensor $\Vert g \Vert$ of the free energy fluctuations as the function of vacuum parameter $\theta$ and temperature $T$, describing fluctuations of the YM vacuum ensemble for a given anomaly factor $c$, decoherence temperature $d$ and critical exponent $e$.} \label{Fdet}
\vspace*{0.5cm}
\end{figure}

Thus, the fluctuating bubble ensemble may intrinsically be examined in terms of the parameters of the underlying large $N$ gauge theory configuration. Moreover, it is evident for a given vacuum bubble system that the principle components of the metric tensor  $\{ F_{\theta \theta},  F_{T T}\}$, which signify self pair correlations, are positive definite functions over a range of the vacuum angle and temperature. Physically, this signifies a set of heat capacities against the intrinsic interactions on the configuration surface $(\mathcal{M}_2(\mathbb{R}),g)$ of the undermining ensemble of parity odd bubbles. The graphical representation of the metric components as the local vacuum correlation functions $\{ F_{\theta \theta}, F_{\theta T}, F_{T T}\}$ as respectively depicted in the Figs.(\ref{Fthetatheta}, \ref{Fthetat}, \ref{Ftt})show that their respective orders of the magnitude are $10^{-2}$, $10^{0}$ and $10^{1}$. 
Furthermore, it is not difficult to see that the determinant of the metric tensor reduces to the following expression
\begin{eqnarray}
|| g ||(x,y) &=& -4 c^2 (2 - e)^2 x^2 (d - y)^{2 - 2 e} \newline \nonumber \\ 
 &+& 2 c (1 - e) (2 - e) (1 + c x^2) (d - y)^{2 - 2 e}
\end{eqnarray}
It is worth mentioning that the parity odd bubble configuration corresponds to a degenerate statistical ensemble for either an absolute value of the vacuum angle $x$ satisfying
\begin{eqnarray}
|x|= \bigg(\frac{e-1}{e-3}\bigg) c^{-1/2}
\end{eqnarray}
or the temperature stays fixed at the decoherence temperature, viz. we have $y = d$, or the critical exponent takes a fixed constant value of $e=2$. Over the domain of the system parameters $\{ x,y \}$, we notice that the Gaussian fluctuations form a stable statistical configuration, as long as the determinant of the metric tensor $|| g ||(x,y)$ remains a positive function on the parametric surface $(\mathcal{M}_2(\mathbb{R}),g)$ as a function of the vacuum angle $x$ and temperature $y$, that's the regions where we have $1< e < 3$ with a given anomaly constant $c>0$. The graphical representation of the determinant of the metric tensor $|| g ||$ as offered in the Fig.(\ref{Fdet}) with respect to fluctuations of $\{x, y \}$ shows a negative amplitude of the order $-0.5$. 

\begin{figure}
\hspace*{1.5cm}
\includegraphics[width=7.0cm,angle=0]{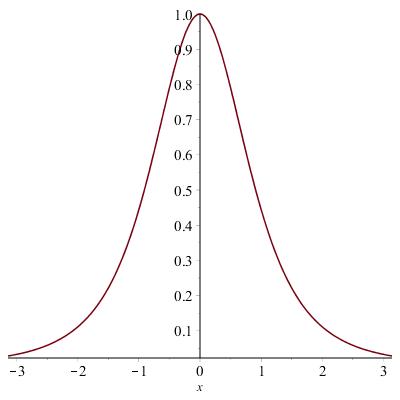}
\caption{The scalar curvature $R$ of the free energy fluctuations as the function of vacuum parameter $\theta$, describing fluctuations of the YM vacuum ensemble for a given anomaly factor $c$ with the critical exponent $e=2$.} \vspace*{0.5cm} \label{Fcurtheta}
\vspace*{0.5cm}
\end{figure}

What follows in this concern is that we specialize ourselves for physically well suited values of the configuration parameters, and subsequently, we have offered the graphical description regarding the stability of parity odd bubble ensembles with the model constant having values of $c=0.5, d=0.197$ and $e= -0.013$ corresponding to the hot QCD deconfining phase \cite{kharjev}.
In order for examining the nature of local transformations concerning the parameters $\{x,y\}$ whose fluctuations form an intrinsic real surface $(\mathcal{M}_2(\mathbb{R}),g)$, we firstly need to determine the functional behavior of the associated Christoffel connection functions. In fact, a direct computation shows that the limiting non-trivial Christoffel connections can be expressed in terms of the third derivative of the free energy. Herewith, we see that the various distinct Christoffel components, which arise in the numerator of the intrinsic scalar curvature \cite{RuppeinerRMP}, reduce to the following expressions
\begin{eqnarray}
F_{111} &=& 0, \newline \nonumber \\ 
F_{112} &=& -2 c (2 - e) (d - y)^{1 - e}, \newline \nonumber \\ 
F_{122} &=& 2 c (1-e) (2-e) x (d-y)^{-e}, \newline \nonumber \\ 
F_{222} &=& (1 - e) (2 - e) e (1 + c x^2) (d - y)^{-(1+ e)}
\end{eqnarray}
Hereby, we observe that the pure vacuum Christoffel component vanishes identically for all values of $x, y \in \mathcal{M}_2$, while the remaining components vanish as well for an ensemble having the critical exponent $e=2$. Further, the pure temperature Christoffel component $F_{222}$ vanishes for the vanishing critical exponent ensembles. It is direct to see that the mixed Christoffel component $F_{122}$ vanishes for both the integer values $e\in \{1, 2\}$. As per this notion of thermodynamic geometry \cite{RuppeinerRMP}, the global nature of phase transition curves can thus be examined over the range of vacuum angle $x$ and QCD temperature $y$ describing a fluctuating ensemble of parity odd bubbles. In particular, we find that the scalar curvature reads as
\begin{eqnarray} \label{freecurvature}
R(x,y) = \frac{k (e-1)(d-y)^{e-2}}{\bigg(1-e+c (e-3) x^2\bigg)^2} 
\end{eqnarray}
We see that the scalar curvature vanishes for the value of the unit critical exponent, viz. $e=1$. It follows further that the scalar curvature vanishes identically at the decoherance temperature $y=d$, unless the system possessed $e=2$. This implies that the decoherance temperature limit $y \rightarrow d$ could also be viewed as the decoupling of statistical interactions or equivalently the formation of thermodynamic equilibrium limit. For the systems having critical exponent $e=2$, we however observe that the scalar curvature becomes independent of the temperature $y$. As a function of the vacuum parameter $x$, we note in this case that the limiting scalar curvature reduces as the following singly peaked squared Lorentzian function
\begin{eqnarray} \label{curvaturee2}
R(x) = \frac{k}{\bigg(1+c x^2\bigg)^2}, 
\end{eqnarray}
as a function of $x$. In this case, we note that the scalar curvature remains regular for all $x, y \in \mathcal{M}_2(\mathbb{R})$, and hence there are no vacuum phase transitions in the limiting configuration of $e=2$. Herewith, from the above Eqn.(\ref{curvaturee2}), we see the corresponding long range global correlation length $l(x,y)$ as the square root of the scalar curvature $R(x,y)$ becomes independent of both the temperature $y$ and decoherence temperature $d$ for the limiting configurations having critical exponent $e=2$. For a given value of the anomaly factor $c=0.5$, the graphical depiction of the above specific scalar curvature $R(x)$ as shown in the Fig.(\ref{Fcurtheta}) in the vacuum angle fluctuation range of $-\pi \le x \le \pi$ demonstrates a Lorentzian type peak of the unit order magnitude. 
Similarly, for an arbitrary ensemble having critical exponent $e=3$, we find that the scalar curvature $R(x, y)$ becomes independent of the vacuum angle $x$. As a function of the system temperature $y$, we observe that it satisfies an equation of straight line of slope $-k/2$ and intercept $kd/2$, viz. we have the following limiting scalar curvature 
\begin{eqnarray} \label{curvaturee3}
R(y) = -\frac{k}{2} y + \frac{kd}{2},
\end{eqnarray}
where $k$ is standard Boltzmann constant. It is worth mentioning further that the slope of the above straight line remains independent of the chosen statistical ensemble, while the intercept depends solely on the decoherence temperature $d$ of the undermining vacuum system. Importantly, we see that the scalar curvature remains regular for all $x, y \in \mathcal{M}_2(\mathbb{R})$, and hence there are no vacuum phase transitions in this limiting case of $e=3$, as well. 
\begin{figure}
\hspace*{1.5cm}
\includegraphics[width=8.0cm,angle=0]{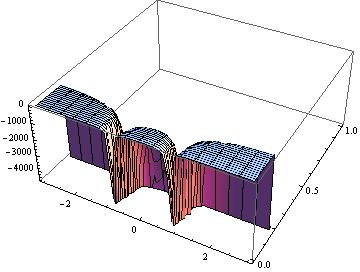}
\caption{The scalar curvature $R$ of the free energy fluctuations as the function of vacuum parameter $\theta$ and temperature $T$, describing fluctuations of the YM vacuum ensemble for a given anomaly factor $c$, decoherence temperature $d$ and critical exponent $e$.} \vspace*{0.5cm} \label{Fcur}
\end{figure}
On the other hand, we find that the intrinsic scalar curvature as given in Eqn.(\ref{freecurvature}) doubly diverges for the following absolute value of the vacuum angle
\begin{equation}  \label{vacuumangle}
| x | = \pm \sqrt{\frac{1}{c} (\frac{e-1}{e-3} )},
\end{equation}
whenever the system possesses a critical exponent $e \not\in \{ 1,2,3\}$ and $y\ne d$, as the phenomenon of decoupling happens at the decoherence temperature. Herefore, under the Gaussian fluctuations, the present examination shows that a typical parity odd bubble hot QCD configuration is globally regular over all possible domains of the parameters $\{x,y\}$ except for the two values of the vacuum angle $x$ as given in the  Eqn.(\ref{vacuumangle}). Further, the graphical depiction of the scalar curvature as represented in the Fig.(\ref{Fcur}) shows that there exist a band structure with a relatively large negative peak of an order of the magnitude $-5 \times 10^{3}$. 

As a matter of the fact, the correlation length of underlying nearly equilibrium vacuum system is globally characterized by the square root of the scalar curvature $R(x, y)$ of the surface $(\mathcal{M}_2(\mathbb{R}),g)$. This follows directly from the fact that the scalar curvature $R(x, y)$, arising from the Gaussian fluctuations of the free energy $F(x, y)$ as a function of the vacuum angle and hot QCD temperature $\{x, y\}$, vanishes identically at the decoherence temperature $y=d$ and diverges for the above mentioned double absolute values of $x$, see in particular the  Eqn.(\ref{vacuumangle}).

Notice further that at the deconfining phase transition point corresponding to the critical exponent $e=-0.013$, the underlying Weinhold geometry of the parity odd bubble vacuum ensemble corresponds to a well-defined statistical basis and is a negatively curved manifold as long as we have $e <1$, and the system remains in the deconfining phase, viz. we have the temperature range $y <d$, except for the configurations having $e=2$. In this case, we observe further that the underlying fluctuation manifold $\mathcal{M}_2$ becomes Ricci flat for the unital value of the critical exponent, viz. $e=1$, wherefore the corresponding vacuum configuration corresponds to a noninteracting statistical system. Moreover, this system generically never diverges, namely under goes a phase transition, for a class of well-defined values of system parameters $\{x,y\}\equiv \{\theta, T\}$, except for those $x$ satisfying the Eqn.(\ref{vacuumangle}) concerning the denominator of the scalar curvature, whose roots determine the undermining global pathologies and statistical singularities. 

This illustrates the statistical structures of an ensemble of parity odd bubbles originating via the free energy fluctuations with an incorporation of anomalous couplings to the undermining mesons. We may further see that the underlying fluctuating configuration pertaining to the large $N$ parity odd bubbles leads to an ill-defined statistical basis at the deconfining temperature $y =d$, as the determinant of the corresponding metric tensor $Hess(F(x,y))$ vanishes identically. Such phenomena may well be expected to happen at this value of the temperature because the mass of the $\eta$ particle vanishes at the confining-deconfining temperature boundary, see \cite{Witten79} for a reasoning via scattering theory.

For the above ensemble of parity odd bubble vacuua with a given anomaly factor $c$, we notice that the Riemann Christoffel curvature tensor $R(x, y)$ in particular possesses a nontrivial global behavior over the entire parametric surface $(\mathcal{M}_2(\mathbb{R}),g)$. Namely, by the present intrinsic geometric analysis having free energy $F(x, y)$ as the real embedding map as in Eqn.(\ref{embeddingf}), we conclude that a parity odd bubble hot QCD vacuum configuration generically corresponds to an interacting regular statistical basis over $(\mathcal{M}_2(\mathbb{R}),g)$ under fluctuations of the system parameters $\{(x, y) \ \vert \ x, y \in \mathcal{M}_2(\mathbb{R})\}$, apart form (i) for a repeated pair of vacuum angles $x$ as given by the Eqn.(\ref{vacuumangle}) where the system goes under vacuum phase transitions, and (ii) for the constant temperature line $y=d$, where $d$ is the decoherence temperature of an arbitrary vacuum system having $e \ne 2$. This is because of the fact that the corresponding scalar curvature $R(x, y)$ as depicted in the Eqn.(\ref{freecurvature}) vanishes identically at this value of the temperature, viz. $y=d$, whereby the associated vacuum ensemble corresponds to a noninteracting statistical system. In fact, we notice also that the same conclusion holds further for an arbitrary vacuum ensemble of unit critical exponent parity odd hot QCD bubbles.
\section{Conclusion and Outlook}
Thermodynamic geometry as introduced by fluctuation theory models \cite{rupp1} renders local and global vacuum phase structures via embedded intrinsic metric, hyper scaling theories \cite{McMillan, Simeoni} and Widom's approach to scaling theory where one defined Widom line as the locus of points in the space of parameters corresponding to the maximum correlation length \cite{widom}. Thereby, as mentioned in the introduction, the state-space geometric characterization has heretofore been investigated for black holes and black branes in various string theory and $M$-theory vacuum configurations \cite{0606084v1,SST, 0801.4087v1,BNTBull,BNTBull08, BNTBullcorr,RotBH,BSBR}. As one of the main subject matter of this paper, this has led us in exploring the vacuum stability properties for an ensemble of large $N_c$ gauge theories as a collection of fluctuating metastable states concerning the parity odd bubbles \cite{kharjev}. Our leitmotif is further supported by observations concerning vacuum $SU(N)$ moduli fluctuations, thereby we find in particular that the statistical correlation functions arise solely due to anomalous metastable states, whereas a general statistical ensemble of vanishing anomaly states parametrizes an ideal noninteracting thermodynamical system.

Furthermore, we observe that the configurations with vanishing anomaly contributions have no global statistical correlation. In other words, the global statistical correlations to party odd hot bubbles arise only because of an existence of the metastable states. It is worth mentioning that metastable states arise only when the anomaly term becomes very small, which implies a strong variation to the topological susceptibility with respect to the system temperature. Moreover, we notice that the statistical system concerning an ensemble of parity odd bubbles undergoes a vacuum phase transition at the chiral symmetry restoration temperature as well as at an absolute value of the vacuum angle $\theta$. Interestingly, the long range global correlation length becomes independent of both the temperature and decoherence temperature for the theories having critical exponent $e=2$. 

The above conclusion holds further for a particular choice of the temperature $T= T_d+1$, where $T_d$ is the decoherence temperature. On the other hand, we find that the unit critical exponent vacuum configuration leads to a noninteracting statistical basis corresponding to a constant mass of $\eta^{\prime}$. Our concerning also leads phenomenological investigations to hot QCD bubbles, see \cite{kharjev} for the maximal isospin violation. This proposal could further be explored towards the stability characterization of a rather technical subject, namely, the formation of stable $\theta$ dependent YM configurations and its relation to statistical stability signatures concerning heavy ion collisions and standard model of physics at LHC \cite{LHC}. 

Our consideration does not stop here, but it further continues. We can indeed extend the above setup into various dimensions of QCD research including nuclear collisions, parity conservation, CP violation and others \cite{KharzeevPisarski,VafaWitten}. It would be interesting to explore the notion of this paper for arbitrary finite $N$ possessing a definite periodicity in $\theta$, by considering for instance the fractional identification: $\theta \sim \theta + 2 \pi r$ with $ 0\ \le \ r \ \le 1$ in connection to the equation of motion of the effective theory. In this paper, we however largely concentrated on the analysis of integral vacuum fluctuations which are $2 \pi$ periodic in $\theta$, viz. $r\in \{ 0, 1\}$. The associated consideration concerning the (in)stability analysis of possible (fractional) vacua arising in the intermediate $\theta$ regime with $0 < r:= p/q < 1$ where $p,\ q \in \mathbb{Z} $ such that $p<q$ and $q \ne 0$, we leave open for a perspective investigation.

In this paper, for the standard cases of $r=0,1$, the fluctuation theory analysis as examined in the foregoing sections $3$ and $4$ renders that an ensemble of parity odd bubbles concerning YM metastable vacua generically leads to an interacting statistical basis, possessing an intriguing set of vacuum pair correlation functions and global phase transition curves. It is worth pointing out further that, for a precise prediction of vacuum statistical correlations, an exact calculation of the undermining $SU(N)$ gauge field theoretic potential $U$ would be essential at a finite $N$ in order to offer complete nonequilibrium statistical stability structures of fluctuating parity odd hot QCD bubble configurations. Concerning a class of finite $N$ such backgrounds, we anticipate that the intrinsic geometric setup would render underlying vacuum stability properties and formation of an equilibrium thermodynamic configuration arising via the thermodynamic limit of an ensemble of meta stable states. We leave this problem open for a future research development. 
\section*{Acknowledgements:} 
This research was supported in part by Amity University Haryana and initiated when B.N.T. was visiting the INFN-Laboratori Nazionali di Frascati, Roma, Italy in June 2014. B.N.T. wishes to express his sincere thanks to Professor Jagdish Rai Luthra and Professor Padmakali Banerjee for their valuable support and encouragements towards the realization of this research.

\end{document}